\documentclass{article}

\selectfont

\usepackage{upgreek}
\usepackage[dvips]{graphicx}

\begin{document}

\title{A contact-less 2-dimensional laser sensor for\\
3-dimensional wire position and tension\\
measurements}

\author{Matthias Prall, V. Hannen, R. J\"ohren, H.W. Ortjohann, M. Reinhardt, \\ and Ch. Weinheimer\\ \\All authors are from\\Institut f\"ur Kernphysik, University of M\"unster, Germany
\\Email: matthias.prall@uni-muenster.de\\ \\M. Reinhardt's present address is:\\ Institut f\"ur Biologische Grenz\-fl\"achen\\Forschungszentrum Karlsruhe, Germany\\ \\
To be published in the april 2010 issue of\\IEEE Transactions on Nuclear Science\\ DOI identifier 10.1109/TNS.2010.2042612}

\maketitle

\newpage

\begin{abstract}

We have developed a contact-less two-dimensional laser sensor which
combines position and tension measurements of wires with a diameter of
order 0.2~mm. The sensor consists of commercially available laser
pointers, lenses, color filters and photodiodes.  In our application
we have used this laser sensor in conjunction with an automated
three-dimensional coordinate measuring machine (CMM). The device
allows for a position measurement of wires in three dimensions with an
accuracy of about $10\,\upmu\textrm{m}$. At the same time the wire
tension can be determined with an accuracy of 0.04~N. The device is
operated at a distance of 150~mm from the wire.

For each position measurement, the laser sensor is moved by the
automated CMM into a plane, where the coordinates at which the wires
intersect with this plane are determined. The position of the plane
itself (the third coordinate) is given by the third axis of the CMM
which is perpendicular to this plane.
The precision of the device was determined using
stainless steel wires with a diameter of 0.2~mm and a tension
of 5~N.  We use the sensor for quality assurance of the
wire electrode modules for the KATRIN neutrino mass experiment. These
modules are comprised of two layers of wires, which are 70~mm apart.  
In general, the device presented here is well suited for the measurement of
any complex wire chamber geometry.

\end{abstract}

\def\thesection{\Roman{section}}

\roman{section}

\section{Introduction}

\begin{figure}[tbp]
\centering
\includegraphics[width=3.5in]{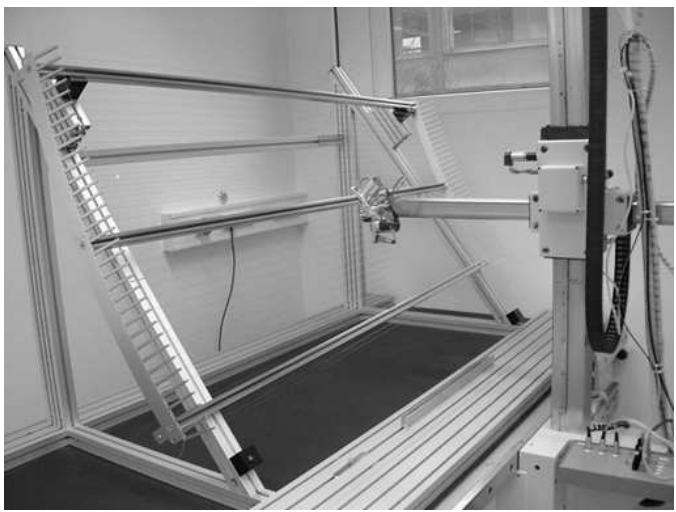}
\caption{Three-dimensional coordinate measuring machine with the laser
  sensor and a \mbox{KATRIN} electrode module with two wire layers in
  our class 10000 clean room \cite{normen} at the University of
  M\"unster. Here, the sensor is being moved upwards during a
  position-measurement.}
\label{fig::maschineMitSensor}
\end{figure}

The main spectrometer of the KATRIN neutrino mass
experiment \cite{Ang04} will be equipped with 248 wire electrodes
(Fig.~\ref{fig::maschineMitSensor}). These wire electrodes consist of
two layers of stainless steel wires, which are approximately 70~mm
apart.  The diameters of the wires are 0.2~mm and 0.3~mm, respectively
for the two layers. We have developed a special laser sensor, which
provides the measurement of the wire positions and their respective
tensions in both layers. The instrument had to meet the following
specifications: 1) Measurement from a distance of 150~mm to avoid the
necessity to move through a wire layer \mbox{during a measurement,} 2)
position measurement with an accuracy of at least 0.1~mm for wires
with a diameter of 0.2~mm and 0.3~mm, 3) tension measurement with an
accuracy of better than 0.1~N to assure that wire tension is less than
1~N above a critical value, 4) total weight of less than 250~g, and,
5) operation under clean room conditions.

With the ascent of wire chambers in nuclear and particle physics many
methods have been developed to precisely measure the position and tension
of thin wires (\cite{Bha88}, \cite{Cal80}, \cite{Cia05}, \cite{Bal07},
\cite{Car97}, \cite{Has08}).  Based on this experience our developments
focused on the above requirements. 


The basic idea of our method is the use of two laser beams of
different wavelength which allows for a very fast kind of triangulation
of the position of the wire in two dimensions. The lasers and the
corresponding light detectors are precisely moved with a
three-dimensional coordinate measurement machine (CMM) to allow an
overall three-dimensional position determination. When the wire is
excited to oscillate by a puff of clean gas, the Fourier transform of
the reflected light signal is used to determine the wire tension. The
combination of these methods is new and fulfills all the mentioned
requirements. Our device could be well suited for the measurement of a
complex detector which consists of many wires e.g. a large wire
chamber.

Our article is organized in the following way: in Section II, we
describe the setup of our laser sensor. Section III covers the
principle and accuracy of the position measurement with our device.
In Section IV the tension measurement is described. In Section V we
discuss sources of systematic errors. Finally, in Section VI and VII we give a comparison to
other methodes and our conclusions.

\newpage

\section{Set-Up of the Laser Sensor}
\label{sec::setup}

\begin{figure}[htbp]
\centering
\includegraphics[width=3.5in]{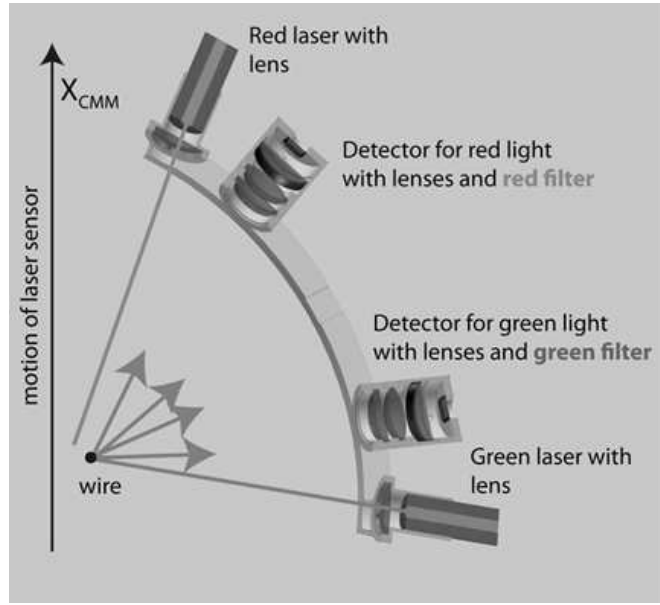}
\caption{Cross-sectional view of the laser sensor. Filters are mounted
  directly in front of the diodes to ensure that the PIN diodes detect
  only red or green light.}
\label{fig::Lasersensor_ganzneu}
\end{figure}

\begin{figure}[t!]
\centering

\includegraphics[width=3.5in]{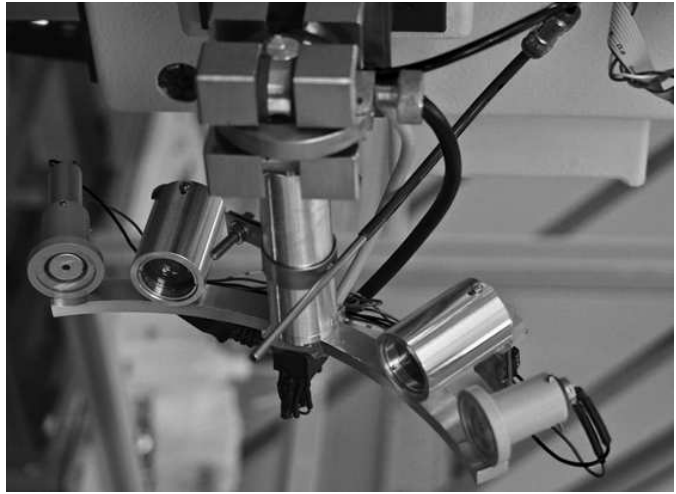}

\caption{The laser sensor with its most important components. A: the
  red laser, B: the detector for red light, C: the gas nozzle
  for the excitation of the wire, D: the detector for green light
  and E: the green laser. }
\label{fig::sensorbild}
\end{figure}

Figure \ref{fig::Lasersensor_ganzneu} shows our contact-less
two-dimensional laser sensor.  The laser beams, which are generated by
ordinary laser pointers with 1~mW power have diameters of approx. 2~mm
at the exit of the laser pointers.  The wavelength of the green laser
is 532~nm. The wavelength of the red laser pointer is in the range of
640~nm to 660~nm.  The battery compartments of the laser pointers were
removed (Fig.~\ref{fig::sensorbild}) and the batteries were replaced
by cables connected to a laboratory power supply
(Fig.~\ref{fig::sensorbild}).  The removal of the batteries also led
to a reduction in deformation of the laser mount due to the weight of
the batteries.  In order to make precise position measurements with
our detector, the laser beams need to have diameters which are small
compared to the diameter of the wire. To meet this requirement we
focus the laser beams with lenses (f=150~mm, diam. 22~mm), where the
focal length corresponds to the desired distance between the laser and
the wire. The intersection point of both laser beams coincides with
the focal points of both lasers.

\begin{figure}[t!]
\centering
\includegraphics[width=3.5in]{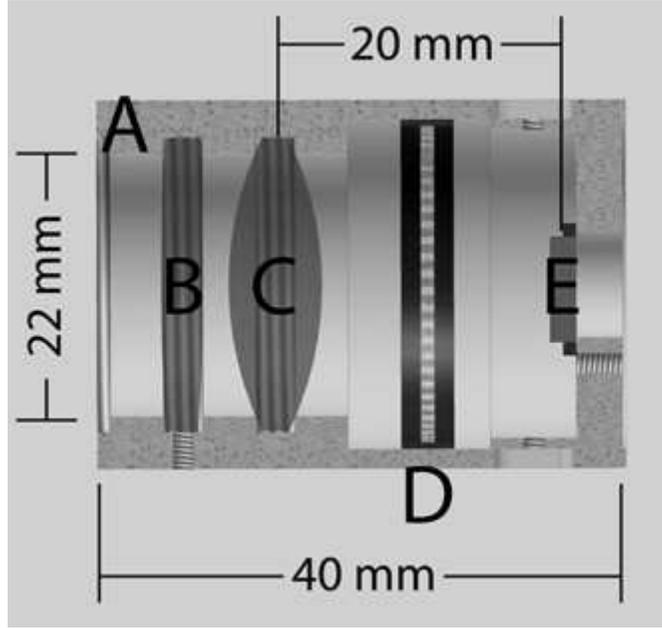}
\caption{Cross-sectional view of a light detector for a reflection. A:
  housing, \mbox{B: lens f = 150~mm,} C: lens f = 20~mm, D: red or green
  filter, E: PIN-diode}
\label{fig::QuerschnittDetektor}
\end{figure}

The light reflected from the wire is detected with PIN diodes.  The
special optics of the light detectors
(Fig.~\ref{fig::QuerschnittDetektor}) has the purpose to focus as much
light as possible onto the PIN diode. The first lens (f = 150~mm)
parallelizes the light, which is reflected from the wire. The second
lens (f = 20~mm) focuses the light onto the PIN diode.  Green and red
bandpass filters \cite{Baader} sit in front of the PIN diodes. 
The light detectors need to be equipped with filters such that they
detect only red or green light. The filters ensure that the device can
be operated under standard light conditions and that crosstalk between
the two different reflection signals can be excluded. The signal of
the PIN diodes is amplified by a custom-made operational amplifier
circuit and recorded by a digitizer card from National
Instruments \cite{NI_card}.

For the position measurement the laser sensor is moved
by a three-dimensional coordinate measurement machine (CMM) across the
wire at a distance of $\approx$150~mm and to detect the laser light,
which is is reflected from a wire
(Fig.~\ref{fig::Lasersensor_ganzneu}).  This distance needs to be
within 5~mm of the focal points due the use of the focusing
optics. Figures \ref{fig::maschineMitSensor} and \ref{fig::sensorbild}
show the sensor mounted on the measurement head of our CMM. Our CMM is
a 20 years old and very robust machine (weight 2500 kg, System C by
Stiefelmayer, Fig.~\ref{fig::maschineMitSensor}).  It has been equipped
 with three motors (type A-Max by Maxon Motor \cite{maxon}) for the
three axes and modified to make it compliant with the operation in
a class 10000 clean room \cite{normen}.

Due to the use of a very robust CMM and speed-regulated direct current
motors we could avoid vibrations of the laser sensor which would
otherwise affect the precision of the measurements. The motors are controlled
with LabVIEW via three RS-232 interfaces of a PC,
which also digitizes the signals from the laser sensors. The complete
software was written with LabVIEW, version 8.0 \cite{labview}.  The
three-dimensional position of the measurement arm, which holds the laser
sensors is measured with three magneto-mechanical encoders with a
resolution of 0.01~mm.  The position information of these three
encoders is recorded with LabVIEW via the
measurement card.  The requirement for the laser sensor to be light-weight
results from the maximum load specification of 250~g for a
sensor on the CMM, which does not compromise the specified precision.

\newpage

\section{Position Measurement}
\label{sec::pos}

\begin{figure}[!htbp]
\centering
\includegraphics[width=3.5in]{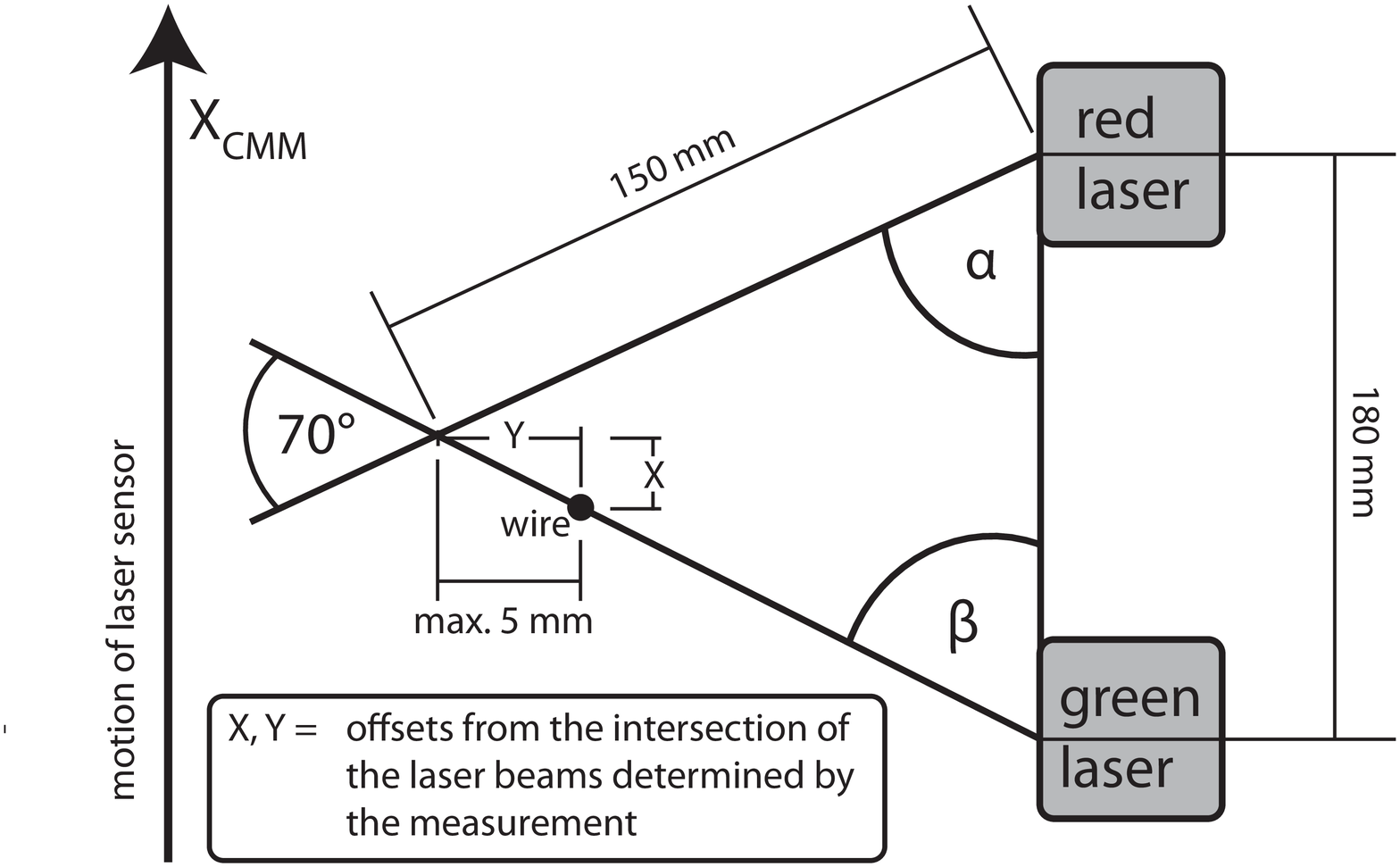}
\caption{Determination of the y-coordinate of the wire.  In our case
  the angle between the two laser beams amounts to
  $70\,^{\circ}=180\,^{\circ} - \alpha - \beta$. This can be
  changed for other applications.}
\label{fig::principle_y}
\end{figure}

In this section we explain the principle of position measurement of a
wire with our laser sensor.  The coordinates x and y of the wire will
be determined by the \mbox{sensor (Fig.~\ref{fig::principle_y})} by
moving it across the wire in x- respectively y-direction and recording
the x(y)-positions $x(y)_{green}$ and $x(y)_{red}$, when the
reflections of the green and red lasers from the wire hit the
detectors.  The z-coordinate is defined by the CMM-position of the
laser sensor. \par

The basic $(x,y,z)$ coordinate system is defined by the CMM encoders.
For a given $z$, a fixed point $(x_{0}, y_{0})$ can be set by setting
the sensor such, that a reference wire on the device to be scanned is
at the intersection of the two laser beams.  The angles of the two
laser beams have to be calibrated beforehand. For this, the sensor has
to be moved across a wire at different y-distances (e.g. 5~mm, 10~mm,
15~mm, 20~mm).  The angles of the laser beams can then be calculated by
fitting the distance versus the position of the maximum of the
reflection in CMM coordinates $x_{green}$ ($x_{red}$).

The position $(x_{wire}, y_{wire})$ of any other wire can be
calculated using this calibration point and the inclinations of the
laser beams $a_{green}$ and $a_{red}$ ($\delta x / \delta y = a$),
where $a_{green} = -1 /\tan{\beta}$ and $a_{red}=1/\tan{\alpha}$
(cf. Fig.~\ref{fig::principle_y}), as follows:

\begin{equation}
\begin{array}{lclll}
y_{wire}  & = & y_{0} & - &  \frac{x_{red}-x_{green}}{a_{red}-a_{green}} \\
  & & \\
x_{wire} &  = & x_{0} & - &  \frac{ a_{red} \cdot x_{green} }{ a_{red} -a_{green}  } \\
\end{array}
\label{eq::positionsformel}
\end{equation}

\newpage

\begin{figure}[!htbp]
\centering
\includegraphics[width=3.5in]{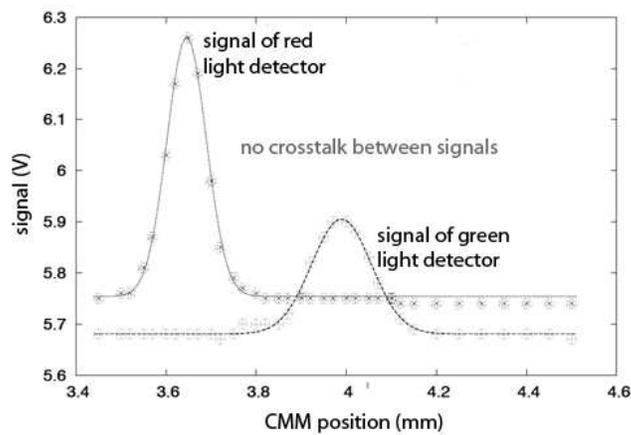}

\caption{Reflections of the red and green laser as measured with the laser sensor.
The position information is derived from the CMM, which moves the sensor across the wire. }
\label{fig::gruener_fit}
\end{figure}


\newpage

In order to determine the accuracy of the position measurement with the laser sensor, we have
mounted a test wire with a diameter of 0.2~mm in a set-up which could
be moved in both x- and y-direction by means of micrometer screws. 
Table \ref{tab::accuracy} shows the results of these tests. 
Positions with index $wire$ were set
using the micrometer screws. Positions with index $meas$ were determined
with the laser sensor. The resolution of the CMM which
moves the sensor is 0.01~mm. To estimate the precision
of the laser sensor, we assume that the relation
between the wire positions and the measured positions can be described
by the equations:

\begin{equation}
x_{meas} = b_{x} + s_{x} \cdot x_{wire} \qquad y_{meas} = b_{y} + s_{y} \cdot y_{wire}.
\label{eq::fitformula}
\end{equation}

\begin{table}[!ht ]
\caption{Settings used for the determination of the accuracy of the
  position measurements with our laser sensor.  Two independent series
  of measurement were taken for x and y, respectively.  Only one
  coordinate was varied for each set of measurements.}
\begin{center}
\begin{tabular}{cc|cc}
$x_{wire}(mm) $ & $x_{meas} (mm)$ & $y_{wire} (mm) $ & $y_{meas} (mm)$ \\
\hline\hline
0.00 & 0.09 & 0.30 & 0.34 \\
0.10 & 0.20 & 0.40 & 0.42 \\
0.20 & 0.30 & 0.50 & 0.54  \\
0.30 & 0.39 & 0.60 & 0.62 \\
0.40 & 0.50 & 0.70 & 0.72 \\
0.50 & 0.60& 0.80 & 0.82 \\
0.60 & 0.69 & 0.90 & 0.92 \\
0.70 & 0.79 & 1.00 & 1.02 \\
0.80 & 0.89 & 1.10 &  1.11 \\
0.90 & 0.99 &          &         \\
1.00 & 1.09 &          &         \\
\hline
\end{tabular}
\end{center}
\label{tab::accuracy}
\end{table}

There can be arbitrary offsets $b_{x}$ and $b_{y}$ between the laser
sensor data and the actual positions. The offsets result from the
initial position of the micrometer screw which moved the test wire.
The slopes $s_{x}$ and $s_{y}$ indicate the size of the systematic error.

Equations (\ref{eq::fitformula}) were fitted to the data (Table
\ref{tab::accuracy}) by a least-squares fit under the assumption of a
constant uncertainty $\sigma_x$ ($\sigma_y$) for the x- and
y-determination by the laser sensor. This resulted in uncertainties of
$\sigma_x = 0.005$~mm and $\sigma_y=0.007$~mm, respectively.

The uncertainty of the measurement of both coordinates is smaller than
the smallest encoder step $s_{min} = 0.01$~mm of the CMM.  This is
reasonable as the statistical limit $L$ for the statistically
distributed difference \mbox{$|x_{wire}-x_{meas}|$} is

\begin{equation}
L = s_{min} \frac{1}{\sqrt{12}} = 0.003 \;\textrm{mm}.
\end{equation}

\begin{table}[!ht ]
\caption{Best fit values for the test of the precision of the 
position measurement}
\begin{center}
\begin{tabular}{lcc}
 & Fit of X values &  Fit of Y values \\
 \hline\hline
offset $b$ (mm) & 0.097 $\pm$  0.003 &  0.042 $\pm$ 0.007 \\
slope $s$ (mm/mm) & 0.993  $\pm$ 0.005 & 0.973 $\pm$  0.009 \\
\hline
\end{tabular}
\end{center}
\label{tab::bestfitvalues}
\end{table}

The results in Table \ref{tab::bestfitvalues} show the precision of
the position measurements at a distance of less than 1~mm from
the intersection of the two laser beams.  Especially the slope of $s_y =
0.973 \pm 0.009$~mm/mm could be due to scale errors of the micrometer
screw. However, it is also possible that the deviation of $s_y$ from unity
points towards a systematic uncertainty in the determination of the position of
the wire far away from the intersection of the laser beams. This shows
that the distance between the wire and the intersection point has to
be minimized in order to guarantee a precision of about
$10\upmu\textrm{m}$.

\section{Tension Measurement}
\label{sec::tension}

When a wire of length $l$, density $\rho$ and cross-sectional area $A$ is
stretched by a force $F$ between two fixed ends, is excited in
transverse direction, it will oscillate.  The fundamental mode has a
wavelength of $\lambda=2\cdot l$. The speed of sound of a wave on such a
wire is given by

\begin{equation}
c= \sqrt{\frac{F}{A\rho}}.
\end{equation}

With $c=\lambda f$, the oscillation frequency $f$ of the fundamental
mode of the wire is given by:

\begin{equation}
f =\frac{1}{2\cdot l}  \sqrt{  \frac{F}{A\cdot\rho}   }
\label{eq::oszigleichung}
\end{equation}

For small oscillation amplitudes this formula is correct to good
approximation, since corrections for the elasticity of the material
can be neglected.

We use Eq.~(\ref{eq::oszigleichung}) to determine the tension of the
wires in our electrode modules with our laser sensor.  The design
value is 10~N for wires with 0.3~mm diameter and 1.8~m length. The
material of the wires is stainless steel \mbox{type 1.4404 ($\rho=
  7980 $\,kg/m$^3$)}.  Thus the oscillation frequency is expected to
be around 37~Hz. \par The frequency is determined through measurement
of the time-dependent reflection amplitude of one of the laser beams,
e.g. the red one.  Note that the wire will reflect light twice per
oscillation period, thus the reflected light pattern has a frequency
$f'=2\cdot f$. We always observed the lowest frequency at this
frequency. Higher harmonics always appear at multiples of $f'$. We
observed that these always exist in our measurements. The first
harmonic is often dominant directly after the excitation and dies out
quickly.  The wire has to be close to the intersection of the two
laser beams, because the diameter of the laser beam needs to be small.
The wire starts to oscillate after it was excited by a short puff of
clean gas (Ar 4.6).  A single pulse is sufficient since even
a small displacement of the wire results in a clearly detectable signal.  The
nozzle, which is used to direct that puff at the wire is labeled with 'C' in
Fig.~\ref{fig::sensorbild}. The nozzle consists of a metal capillary
with an inner diameter of 2~mm. An electronically controlled valve
opens the connection between the nozzle and a reservoir of argon with
a pressure of 8~bar for a duration of 10~ms. This results in a puff which has a diameter of
approx. 10~mm at the location of the wire. The distance which the
argon has to travel between the valve and the exit of the nozzle is
approximately 130~mm (cf. Fig.~\ref{fig::maschineMitSensor} and
  Fig.~\ref{fig::sensorbild}). One can also excite wire oscillations
with a mechanical device, e.g. a small hammer, without damaging the
wires. However, our method is entirely contact-less. The versatility
of the CMM avoids the necessity to adapt any mechanics for different
 module geometries.

\begin{figure}[!htbp]
\centering
\includegraphics[width=3.5in]{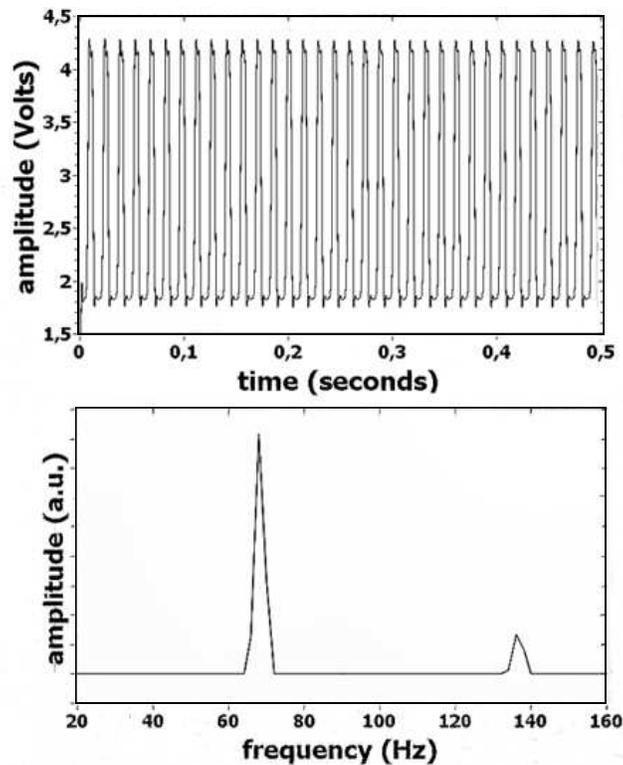}
\caption{Upper picture: Reflected light intensity versus time in
  seconds (the data have an arbitrary offset).  Lower picture: Fourier
  transformation of
  the reflected intensity.}
\label{fig::frequenzbilder}
\end{figure}

The reflected intensity versus time as recorded by our data
acquisition system is plotted in the upper plot of
\mbox{Fig.~\ref{fig::frequenzbilder}}. Also shown, in the lower plot,
is the corresponding Fourier transformation derived with the LabVIEW
function 'Power Spectrum'.  We use a peak finding algorithm (LabVIEW
function 'Harmonic Distortion Analyzer') to determine the position of the
fundamental mode, which is able to interpolate in
steps of 2~Hz resulting from the sampling frequency of 8000~Hz and the
sampling length of 0.5~s.  The measurement stops, when the peak finding
algorithm has detected a frequency within a certain range. In our case,
this range was chosen from 40~Hz to 90~Hz to find frequencies around $f' \approx 2 \cdot 37$~Hz.
The exact values to be
chosen as upper and lower bounds depend on the
variations incurred in the actual wire tensions.\par
  
Our data acquisition system samples the signal from the detector
with $f_{0} = 8000$ samples per second corresponding to a
Nyquist frequency of 4000~Hz. 
Typically, the oscillation of the reflected laser light is
stationary after 1~s and dies out within less then ten
seconds. Therefore, we have chosen to
sample the signal for a few seconds and calculate the Fourier transform every 0.5~s i.e.
every 4000 samples (Fig.~\ref{fig::frequenzbilder} bottom). \par  
  
In order to suppress high frequency noise, we filter the signal with a
third order Butterworth low-pass with a cutoff frequency of 100~Hz in
LabVIEW before the power spectrum of the signal is being computed.
The cutoff frequency is also chosen such that harmonics above the
signal frequency of the fundamental mode are suppressed.  

The choice of this particular filter has no special reason, but it satisfies our requirements nicely.
 Using a software filter from LabVIEW`s library instead of a hardware
 filter maintained our flexibility for different edge frequencies,
 since we commissioned the laser sensor while the first electrode
 modules for KATRIN were built and the design of the last modules was
still undefined.

\begin{figure}[!htbp]
\centering
\includegraphics[width=3.5in]{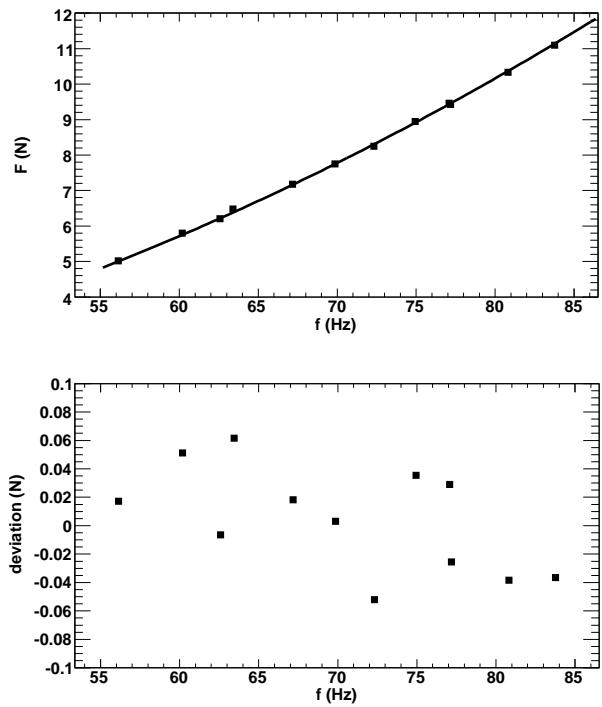}
\caption{Parabolic fit to the measured wire tensions (top) assuming a constant uncertainty
of $\Delta F = 0.04$~N and the corresponding deviations from the parabolic least-squares fit (bottom).}
\label{fig::tension}
\end{figure}

In order to determine the precision of the tension measurement, we
varied the tension of a 176~cm long test wire with 0.3~mm diameter and
measured the frequency with our laser sensor
(Fig.~\ref{fig::tension}).  The tension was measured independently
with a calibrated load cell (WRC-01 USB, model DBBP-20 by Weiss
Robotics).  Following Eq.~\ref{eq::oszigleichung}, we fit the
resulting data with a parabola

\begin{equation}
F(f) = a \cdot f'^2,
\end{equation}

where $a$ is a free parameter. As mentioned above the measured
frequency $f'$ corresponds to twice the oscillation frequency of the wire $f$.
Assuming a constant uncertainty of the force $\sigma_F$
the least-squares fit results in $\sigma_F = 0.04$~N for the
uncertainty of the tension determination (Fig.~\ref{fig::tension}).
The measured deviations from the fit (Fig.~\ref{fig::tension} bottom) confirm the
precision of the method.

The fit result \mbox{$a =(1.587 \pm 0.0002) \cdot
10^{-3}~\textrm{N}/\textrm{Hz}^2$} can be compared to the expectation
from \mbox{Eq. (\ref{eq::oszigleichung}).} \par 
One has to keep in mind that there are uncertainties of the parameters $l$, $A$ and $\rho$. 
The reflection of the standing wave at the ends of the wire is not necessarily ideal.
We do not know the uncertainty of \mbox{$\rho=7980 \,\textrm{kg}/\textrm{m}^3$} but could measure
$l$ and the wire diameter $d$ directly.\par 
The measured parameters
(\mbox{$l=1.76\pm 0.003$ m} and \mbox{$ d = 0.29\pm 0.005$~mm}) imply a theoretical value \mbox{$a=(1.633\pm 0.057)\cdot 10^{-3}\,\textrm{N}/\textrm{Hz}^2$.} 
I.e. the theo\-retical prediction is unsignificantly larger than the observation.

\newpage

\section{Error Sources and Limitations}

For the stainless steel wires under scrutiny (diam. 0.2~mm - 0.3~mm,
$l\approx 1.8$ m, $\rho=7980\, \textrm{Kg}/\textrm{m}^{3}$), we
observe that the tension can no longer be determined when it is less
than $\approx$ 1~N. Around that value, the intensity variation of the
reflected laser light is no longer periodical. \\ An important
specification for our sensor is a maximum weight of 250 g.  The weight
of the sensor causes the arm of the CMM to sag by 30~$\upmu$m when
fully extended (1~m). This sag is tolerable for us, and can be
compensated for.\par 
The precision of the position measurements
depends on the correct alignment of the two laser beams. When the
laser beams are not perfectly aligned with the measurement plane, the
beams hit the wire a slightly too early or too late, when the sensor
moves across a wire. This leads to an error in the determination of
$x_{green}$ and $x_{red}$ leading to a systematic error of the wire
position.\par 
Suppose that the red laser is correctly aligned, but the
green laser is displaced by 0.5~mm with respect to the measurement
plane. Let the measured wire lie in the (x,z)-plane and both angles
$\alpha$ and $\beta$ are $35^{\circ}$.  Let us further assume,
that the measured wire has an inclination of $5^{\circ}$ with respect
to the z-axis.  The resulting errors due to this misalignment will be:

\begin{equation}
\begin{array}{l}
|\Delta x_{wire}| = |\frac{\partial x_{wire}}{\partial x_{green}}\Delta x_{green} | = 0.02 \,\textrm{mm} \\
\\
|\Delta y_{wire}| = |\frac{\partial y_{wire}}{\partial x_{green}}\Delta x_{green} | = 0.02 \,\textrm{mm} \\
\end{array}
\end{equation}

\section{Comparison with other Methods}

There are numerous methods which can be used to measure wire tensions
and positions from a distance. In a well-known method for measuring
the wire tension, an external magnetic field is applied and mechanical
oscillations of the wire are induced by passing an alternating current
through it. As long as no special techniques are employed one has to
scan a frequency range in order to find the resonance frequency of the
wire, which provides the tension measurement. This process can take
several minutes \cite{Bha88}. This time can be drastically shortened
using a sudden excitation of the wire in the magnetic field with a
current pulse \cite{Cal80} or an electrostatic excitation through
capacitive coupling \cite{Cia05}. In any case, one still needs to
connect the wires to an external circuit, which is still
time-consuming. Moreover, it could lead to a damage of the wires.

It has also been reported that one can excite wire oscillations with a small
hammer, even if the wire thickness is only 30 $\upmu$m, without
causing any damage \cite{Bal07}. It could be difficult to adapt the
corresponding hardware to changing geometries. Thus the real advantage
of our excitation method is not that we avoid damage, but that we
avoid the necessity to adapt mechanics to changing module
geometries.\par

Other authors describe wire position measurements with a laser diode
with an accuracy of better than $10\,\upmu$m for relative wire
separations of thin wires ($<76\,\upmu$m diameter) in a wire chamber
but with a principle which is restricted to \mbox{one dimension
  \cite{Car97}.} There is also an example \cite{Has08} of a
measurement with two CCDs in 2 dimensions with an accuracy of
$O(1)\,\upmu$m but from a smaller distance \mbox{of $\le 40$ mm.} The
last method is the most precise method in two dimensions of which we
are aware, however, it cannot be used to measure wire tensions.

\section{Conclusion}
\label{sec::additional}

We developed a new two-dimensional laser sensor for the measurement of
parameters of thin wires. The key feature of our novel sensor is the
combination of a two-dimensional position measurement with a tension
measurement at a distance of 150~mm with respect to the wires. Both
the position and the tension measurements can be performed within a few seconds
without touching the wire.\par 
The laser sensor is moved by a
precise three-dimensional coordinate table and the intensity of the
reflected light is recorded. With this a position resolution of better than
$10\,\upmu$m has been achieved.  For the measurement of the wire
tension,
oscillations are excited by a
puff of gas blown at the wire, and the intensity variation of the reflected light is
recorded. The fundamental frequency of this variation is found by
means of a fast
Fourier transform and yields the wire tension with a precision
of 0.04~N.  An important property for our application is the
light-weight construction of the laser sensor ($< 250$~g).  We apply
this wire measurement system for the quality assurance of the wire
electrode system of the KATRIN neutrino mass experiment
\cite{Ang04}. In a second version of the sensor we have now enlarged the
detector lenses to a diameter of 40~mm in order to increase the angle
of acceptance to $\pm 10^\circ$ to measure wires non-perpendicular to
the laser plane.  \par 
In practice, our sensor needs $\approx\,20 $
minutes to measure the tensions of all 120 wires of a module and
$\approx 30$ minutes to scan 120 wire position in one plane. A large
fraction of the time, our CMM has to move the sensor from one wire to
another. As soon as the CMM has positioned the sensor on a wire, it
takes between two and four seconds to obtain a result for a wire
tension.  A position measurement, i.e., a scan across a wire, takes
$\approx$ 10~s.  If it is required to speed up the measurement process, a
faster CMM machine could be used, up to the limit of vibration damping
of the sensor.


%

\section*{Acknowledgment}
The authors would like to thank A. Gebel for his help as well as the mechanics and electronics workshops of our
institute for their support.  We would like to thank H. Gottschlag
(ALICE Collaboration), who has provided us with a 1-dimensional sensor
for prototype experiments. This sensor was developed for the
measurement of wire tensions of TRD readout chambers of the ALICE
experiment \cite{GSI03}.  He also provided support for the first
measurements, which evolved into the idea for our new sensor.\par
We also thank J.P. Wessels for carefully reading and correcting the article.\\ This
project is funded by the German Ministry of Research and Education
(BMBF).

\end{document}